\documentclass[11p]{article}

\usepackage[authoryear,longnamesfirst]{natbib}
\usepackage[T1]{fontenc}
\usepackage[english]{babel}

\usepackage{xfrac,changebar}
\usepackage{color}
\definecolor{darkgreen}{rgb}{0.0,0.7,0.0}

\definecolor{darkred}{rgb}{0.75,0.0,0.0}

\usepackage{amsmath,amsthm,amssymb,amsfonts}
\usepackage{subfig}
\usepackage{etoolbox}
\newcommand{\TT}{\mathcal{T}}
\newcommand{\N}{\mathbb{N}}
\newcommand{\R}{\mathbb{R}}
\newtheorem{lem}{Lemma}
\newtheorem{cor}{Corollary}
\newtheorem{thm}{Theorem}

\theoremstyle{definition}
\newtheorem{defn}{Definition}

\theoremstyle{remark}
\newtheorem{rem}{Remark}

\title{Optimality and sustainability of hybrid limit cycles in the pollution control problem with regime shifts}  
\author{Dmitry Gromov${}^1$, Tadashi Shigoka$^{2}$, Anton Bondarev$^{3}$\\[15pt]
${}^1$ Department of Mathematics, University of Latvia,\\ Rai\c{n}a bulv\={a}ris 19, R\={\i}ga, LV-1586, Latvia\\[5pt]
${}^2$ Kyoto Institute of Economic Research,\\ Yoshida-honmachi, Sakyo-ku, Kyoto, 606-8501, Japan\\[5pt]
${}^3$ International Business School Suzhou,\\ Xi'an Jiaotong-Liverpool University,\\ Ren'ai Road 111, 215123, Suzhou, P.~R.~China}
\date{}

\begin{document}

\begin{titlepage}
\maketitle
\end{titlepage}

\begin{abstract}
In this paper, we consider the problem of pollution control in a system that undergoes regular regime shifts. We first show that the optimal policy of pollution abatement is periodic as well, and is described by the unique hybrid limit cycle. We next introduce the notion of an environmentally sustainable solution, and demonstrate that such a policy is the only one that yields the best possible trade-off between steadily achieving profit and ensuring environmental preservation. In contrast to that, the policy that is not environmentally sustainable eventually enters stagnation. To further illustrate our findings, we compare the optimal periodic solution with a myopic one. Interestingly enough, the myopic solution yields higher overall payoff in the short-run, but completely fails in the long-run, while the environmentally sustainable policy yields maximal payoff and preserves the environment over the infinite time interval.\smallskip

\end{abstract}



\section{Introduction}

There is a growing interest to hybrid control problems and piece-wise smooth systems in environmental and economic literature. This is because many environ\-mental-economic phenomena naturally admit a hybrid description with multiple regimes of dynamics. The recent contributions to the field are \citet{LKN:2001,Hoekstra/Bergh:2005, Belyakov/Veliov:2014,BDV:2015,ZLS:2017,Pichika/Zawka:2018,MPF:2019,Pichika/Zawka:2019} among others. Some of these papers also considered cyclic or fluctuating optimal dynamics. However, to the best of our knowledge, the optimal policy in the form of hybrid limit cycle (HLC) was first reported in \citep{GBG:21OL}. We contribute to the line of research pioneered by that paper by proving the uniqueness of such a cyclical policy and further expanding it within the context of applied environmental problems.

In this paper, we consider a stylized pollution control problem (cf.~\cite{Dockner}, \cite{BZZ:05}) such that the regenerative ability of the environment changes with time. There have been different approaches to modeling the self-cleaning rate of the environment, see, e.g., \cite{Liu:18}. When considering a short term dynamics, it is generally assumed that the self-cleaning rate is negatively affected by the stock of pollution, \cite{Cheve:00,OBG:14}.  However, as we are interested an infinite horizon solution, we consider only the most substantial, seasonal aspect of the absorption rate variability, while neglecting the variation due to the changing stock of pollution. This seems to be a reasonable assumption as long as we are interested in studying normally functioning ecosystems. Thus, we assume that regenerative capacity undergoes seasonal (periodic) variations, which is in line with biological evidence for, e.g., particulate air pollution, \cite{Nieuwenhuijsen:07,Wang:13} or polycyclic aromatic hydrocarbons pollution, \cite{Klamerus:18}.
 	
A particular feature of the considered model is that the canonical system is decoupled in the state and the co-state. This is due to the cost function (ecotax) being linear in the state (pollution level). In the existing literature there are two main approaches regarding the choice of the cost function (also see \cite{Jorg:10} for a detailed overview of different taxation scheme in the context of game-theoretic pollution control). Either the cost function represents the environmental damage that a social planner tries to minimize. In this case, the cost function is typically chosen to be quadratic in the stock of pollution, \cite{MXZ:03}. However, in economically oriented applications the cost function is rather considered to be linear in the stock of pollution, \cite{Luqman:18}. In this work, we adopt the second assumption.

For the formulated problem, it was demonstrated in \cite{GBG:21OL} that any candidate optimal solution of the considered problem eventually converges to an HLC of a special type.
 We prove the existence of a {\em unique} HLC both in the state and the adjoint variables that corresponds to the optimal solution of the considered problem.

Note that there have been several studies of hybrid optimal control problems with a (potentially) infinite number of regime shifts, e.g., \citep{CEMS:07,Schollig:07}, where a class of hybrid systems with regional dynamics and a bounded number of regime switches was considered.
However, no explicit solutions were found.
Somewhat different approaches were pursued in \citep{Savkin:99,matveev2000,ZLS:2017}, who also studied cyclic systems and proved the existence of HLCs, but did not investigate their optimality.
In a recent paper \citep{RG:19} (see also \cite{BG19} and \cite{Seidl:19}), a discounted hybrid optimal control problem (DHOCP) with control switches determined by the state (state-driven switching using the terminology from \cite{Gromov2017}) was studied, and it was shown that such a problem cannot have an HLC as an optimal solution.
In contrast to the mentioned case, we consider a DHOCP with time-driven switching (a system that undergoes regime changes at fixed time instants).

Furthermore, we analyze the dependence of optimal control profiles on the parameters of the system and define a special class of optimal solutions. Specifically, we say that the optimal control is \emph{environmentally sustainable} if it stays within the admissible set and does not take boundary values over the time intervals of non-zero length. To interpret this requirement, we recall that the studied DHOCP models a pollution control problem in which the decision maker (the manufacturer) aims at finding a trade-off between the profit obtained from production and the losses due to the pollution. Here, the amount of the pollution-related losses is determined by the size of the ecotax imposed by the regulator. It turns out that in attempt to reduce the total pollution by increasing the respective fees, the regulator may create the situation when the net profit of the manufacturer turns to zero, thus rendering the production not profitable. The condition of {\em environmental sustainability} characterizes exactly the opposite, i.e., the solutions that lead to the increase in the net profit over the planning horizon (although the instantaneous profit can be negative over short time intervals).

We believe that it is instructive to compare the optimal solutions with the so-called {\em myopic} ones, which are obtained by constraining the instantaneous payoff to be non-negative. In this case, the  accumulated profit over a short planning interval increases at the cost of environmental degradation, but eventually this leads both to a critically polluted environment and a negative profit stream. This scenario clearly demonstrates that the choice of the planning horizon not only makes a difference, but can potentially lead to catastrophic consequences.

We thus claim that a control bounded to lie within the interior of the admissible set is the only one which is both long term profit-maximizing and environmentally sustainable.
 These results continue the line of research aimed at determining the optimal conditions and regulatory mechanisms to mitigate the pollution problem, see, e.g., \cite{Shortle:01,Jouvet:05,Arguedas:17,Arguedas:20}. In this connection, we specifically mention papers \cite{deZeeuw:12,Nkuiya:16}, where the problem of pollution control was considered in the presence of regime shifts.

The rest of this paper is organized as follows.
In Section~\ref{sec:2}, the DHOCP under study is formulated, in Section~\ref{sec:3} we work out in detail the optimal control structure and determine the unique optimal solution.
In Section~\ref{sec:4}, we introduce the notion of an environmentally sustainable optimal control and formulate the conditions on the parameters such that this property is satisfied.
We further compare the environmentally sustainable solution with the myopic one and demonstrate the long-term advantage of the sustainable solution.
Section~\ref{sec:5} concludes the paper.

\section{Problem statement}\label{sec:2}

\begin{subequations}\label{eq:problem0}
We consider the problem of controlling the emission of pollutants \cite[Sec.~5]{Dockner} such that the dynamics of the stock of pollution are governed by the linear differential equation
\begin{align}
	\dot{z}=\xi v -\delta z,\quad z(0)=z_0\ge 0,\label{eq:sys0}
\end{align}
where $z$ is the stock of pollution within a fixed natural reservoir (e.g., a lake), $v\ge 0$ is the emission rate, $\xi\in(0,1)$ is the fraction of the emitted harmful substance that gets accumulated in the reservoir, and $\delta>0$ is the rate of self-cleaning. Obviously, since the control $v$ is non-negative, the stock of pollution is non-negative as well, i.e., $z(t)\ge 0$ for all $t\ge 0$.
 
The profit function $P(v)=av(b-v/2)$ is a concave function of the pollution rate $v$, and has the property of decreasing marginal returns for all $v(t)\in[0,b]$, where $b$ is the maximal admissible emission rate.
While the expression $v(b-v/2)$ describes the rate of production, the coefficient $a$ is used to transform the flow of production to the flow of profit.
Note that within the admissible interval, the rate of pollution is uniquely related to the rate of production.

On the other hand, the term $qz$, where $q$ is a positive constant, corresponds to the fines the agent has to bear (e.g., an ecotax). In this formulation, we follow the common rule of thumb that ``an environmental tax generally should be levied as directly as possible on the pollutant or action causing the environmental damage'', \cite{oecd:11}. We assume for simplicity that there is a single source of pollution, so the polluting agent say, an enterprise, is responsible for the whole amount of pollution.
The stock of pollution $z(t)$ increases from the productive efforts of the agent, associated with $v(t)$, whereas a fraction $\xi$ of those efforts is translated into an increase in the pollution stock via \eqref{eq:sys0}.
The latter decays over time at the rate $\delta$.

According to the above, the payoff functional is defined as the discounted net profit $L(v,z)=P(v)-qz$ obtained by the agent in the problem with infinite time horizon:
\begin{align}
	\bar{J}(z_0):=&{ }\max_{v(\cdot)}\int_{0}^{\infty}\mathrm{e}^{-rt}\left[av(t)\left(b-\frac12v(t)\right) - q z(t)\right]dt,\label{eq:J0}
\end{align}\end{subequations}
where $r>0$ is the discount rate, and the control input satisfies $v(t)\in[0,b]$ for all $t\ge 0$.
Furthermore, we have that for all $z_0\ge 0$, the state $z(t)$ is non-negative.


The problem \eqref{eq:problem0} is normalized by introducing new state and control variables:
\[u(t)=\frac{1}{b}v(t),\quad x(t)=\frac{q}{ab^2}z(t).\]
Note that the normalized control takes values in the interval $[0,1]$.
Both the control $u(t)$ and the state $x(t)$ are now dimensionless.
 After the normalization, the optimal control problem \eqref{eq:problem0} can be written as follows:
\begin{subequations}\label{eq:problem}\begin{align}
	J(x_0):=&{ }\max_{u(\cdot)}\int_{0}^{\infty}\mathrm{e}^{-rt}\left[u(t)\left(1-\frac12u(t)\right) - x(t)\right]dt\label{eq:J}\\
	&\dot{x}=\beta u -\delta x,\quad x(0)=x_0.\label{eq:sys}
\end{align}\end{subequations}
Here, $\beta=\frac{\xi q}{ab}>0$, the state $x(t)$ is non-negative, and the control input $u(t)\in[0,1]$ for all $t\ge 0$.
 Let us take a closer look at $\beta$.
It is the ratio of the penalty related component $\xi q$ to the production related (linear) component $ab$.
To put it differently, if $\beta>1$, and self-cleaning can be neglected, a small amount of the emitted pollutant contributes more to the penalty than to the profit.

We further make the key assumption of the paper in that we postulate that the self-cleaning rate $\delta$ in \eqref{eq:sys} changes as a function of time.
Specifically, we assume that the whole time interval $\TT=[0,\infty)$ is divided into an infinite number of equal intervals of length $T$, whereas each of these intervals is subdivided into two parts: $[k,k+\alpha)T$ and $[k+\alpha,k+1)T$, where $\alpha\in (0,1)$ and $k\in\mathbb{N}_0$.
During the first subinterval, the system is in the first mode, $\delta=\delta_1> 0$, while during the second subinterval the system is in the second mode, $\delta=\delta_2> 0$.
To put this formally, we define
\begin{align}\label{eq.delta}
\delta(t):=\begin{cases}\delta_1> 0,& t\in[k T,k T+\alpha T),\\\delta_2> 0,& t\in[kT+\alpha T,(k+1)T).\end{cases}
\end{align}
Our goal is to understand the behavior of the solution to \eqref{eq:J} and \eqref{eq:sys} given the piecewise character of \eqref{eq.delta}. It is assumed that $\delta_1\neq \delta_2$ as otherwise the problem becomes trivial.

The problem \eqref{eq:J}-\eqref{eq.delta} corresponds to the important problem of pollution management: the agent aims at maximizing the profit stream associated with production, resp., polluting activity $u$.
The payoff increases with the production of the pollutant, but at the same time decreases with the deterioration of the environment.
The pollution stock increases from production, but can  regenerate itself at the rate $\delta$.
For many environmental processes, the regeneration rate varies periodically over time (e.g., over the summer and winter periods), so such a formulation of the DHOCP appears natural.

\section{The optimal control}\label{sec:3}
In this section, we improve the result of \cite{GBG:21OL} by proving that the optimal control to \eqref{eq:problem} is uniquely determined and forms a hybrid limit cycle. The canonical system for \eqref{eq:problem} has the following form:
\begin{subequations}\label{eq:CS}
\begin{equation}\label{eq:x-inf}
\dot{x}(t):=\begin{cases}\beta u^*(t)-\delta_1 x(t),\hspace*{65pt}& t\in[k T,k T+\alpha T),\\
\beta u^*(t)-\delta_2 x(t),& t\in[kT+\alpha T,(k+1)T),\end{cases}
\end{equation}
\begin{equation}\label{eq:lambda-inf}
\dot\lambda(t):=\begin{cases}(\delta_1+r)\lambda(t)+1=\rho_1\lambda(t)+1,& t\in[k T,k T+\alpha T),\\
(\delta_2+r)\lambda(t)+1=\rho_2\lambda(t)+1,& t\in[kT+\alpha T,(k+1)T),\end{cases}
\end{equation}
\end{subequations}
where $k\in \N$ and $\rho_i=r+\delta_i$, $i=1,2$. Later on, we will consider the case $T=1$. This does not reduce the generality of the result, as the duration of the interval can be always set to $1$ be a proper scaling of time. For this case, we will denote $t_s=\alpha T=\alpha$ and call $t_s\in(0,1)$ the {\em switching time}. 

The optimal control is given by
\begin{equation}\label{eq:u-opt}
	u^*(t)=\begin{cases}
		0,&\lambda(t)<-1/\beta,\\
		\beta\lambda+1,&\lambda(t)\in [-1/\beta, 0],\\
		1,&\lambda(t)>0.\end{cases}
\end{equation} 
Note that $\lambda$ never turns positive, hence the last case can be discarded.

First, we observe that the state variable is bounded.
\begin{lem}\label{lem1}For any initial value $x_0\ge 0$ and $\lambda_0\in \R$, the state variable $x(t)$ corresponding to \eqref{eq:CS} and \eqref{eq:u-opt}, satisfies 
\[0\le x(t) \le\max\left(x_0,\frac{\beta}{\min(\delta_1,\delta_2)}\right),\quad t\ge 0.\]
\end{lem}
\begin{proof}Note that the right-hand side of \eqref{eq:x-inf} is negative for all $u\in  [0,1]$ and all $x>\frac{\beta}{\min(\delta_1,\delta_2)}$ and non-positive for $x=\frac{\beta}{\min(\delta_1,\delta_2)}$. Thus, the required result follows.
\end{proof}

This fact, along with the standard arguments from infinite horizon optimal control theory, yields the following characterization of the optimal solution.
\begin{thm}\label{thm1}The solution to \eqref{eq:CS} and \eqref{eq:u-opt} satisfying $(x(0),\lambda(0))=(x_0,\lambda_{eq})$ with
\begin{equation}\label{eq:lambda-eq}
\lambda_{eq}=\frac{\rho_ 1-\rho_2+\rho_2 e^{\rho_ 1 t_s}-\rho_ 1 e^{\rho_2 \left(t_s-1\right)}}{\rho_ 1 \rho_2 e^{\rho_2 \left(t_s-1\right)}-\rho_ 1 \rho_2 e^{\rho_ 1 t_s}},
\end{equation} 
is the unique optimal solution to \eqref{eq:problem}, \eqref{eq.delta}.
\end{thm}
\begin{proof}See Appendix.
\end{proof}
Let $(x^{\ast }(t),\lambda ^{\ast }(t),u^{\ast }(t))$\ be the solution to
(4) and (5) with $(x^{\ast }(0),\lambda ^{\ast }(0))=(x_{0},\lambda _{eq})$
so that $\{(x^{\ast }(t),\lambda ^{\ast }(t),u^{\ast }(t))\}_{t\geq 0}$\
constitutes the unique optimal solution to (2), (3) by Theorem 1. Finally, we
show that the optimal solution $x^{\ast }(t)$\ converges to a hybrid limit
cycle $x_{h}(t)$\ such that $x_{h}(t)=x_{h}(t+T)$\ for each $t\in \mathbb{R}%
_{+}$\ and that $x_{h}(s)\neq x_{h}(0)=x_{h}(T)$ for each $s\in (0,T)$.

\begin{thm}\label{thm2}For any $x^{\ast }(0)=x_{0}\geq 0$, the optimal
solution $x^{\ast }(t)$\ to (2), (3) asymptotically converges to a uniquely
defined hybrid limit cycle $x_{h}(t)$ as $t\rightarrow \infty $.
\end{thm}
\begin{proof}See Appendix.
\end{proof}
\begin{rem}Note that although the optimal control does not depend on the initial value of the state $x_0$, it can be readily seen that the optimal hybrid limit cycle evolves within the interval $(0,\beta/\min(\delta_1,\delta_2))$. That is to say, for large initial values, the optimal solution consists of two phases. During the first one, the agent does not pollute for certain time until the state approaches the hybrid cycle. Then, the agent switches to the control that drives the system along the hybrid limit cycle. 
\end{rem}

We thus have shown that in contrast with DHOCP with state-driven switches, the one with time-driven switches does not only allow for HLC as optimal behavior, but this HLC is \emph{the only and unique} optimal trajectory. To our knowledge, this is the first result on optimality of such dynamics.
In the next section, we will carry out a qualitative analysis of this solution and study environmental and economic implications of it.

\section{Qualitative analysis of the optimal solution}\label{sec:4}

When considering optimal control problems in economic applications, it is common to postulate that the optimal control does not belong to the boundary of the admissible set except for isolated instants of time.
Below, we will give an economic interpretation of this requirement within the context of our problem and provide conditions for it to hold.

Recall the definition of the optimal control \eqref{eq:u-opt} and note that $u^*(t)$ takes a boundary value if either $\lambda(t)>0$ or $\lambda(t)<-1/\beta$.
It was previously stated that $\lambda(t)$ may never exceed $0$, which implies that it is not optimal for the decision maker to ramp up the production to the maximum level.
On the other hand, it may happen that the value of $\lambda(t)$ decreases below $-1/\beta$ over some time interval.
The respective optimal control has to be set to $0$ over this interval, which implies that the agent halts production and lets nature (partially) recover and lower the current level of pollution.
Before we proceed to the analysis of the implications of such an outcome, we introduce the following notion, which we will employ throughout the rest of this paper.

\begin{defn}
The optimal strategy $u^{\ast}(t)$ is said to be {\em environmentally sustainable} if it does not take on boundary values  except at isolated instants of time, i.e.,
$\lambda(t)\in [-1/\beta,0]$ $\forall t\ge 0$.
\end{defn}
The  motivation behind such a notion is as follows.
The situation when the control falls below the value zero occurs when the loss due to the ecotax, which is proportional to the current level of pollution, exceeds the profit obtained from production.
Should such a situation occur, the manufacturer has to halt production until the level of pollution, and the loss, decreases to be below the level determined by the current production.
The relation of the loss to the profit is determined by the parameter $\beta$. 
Recall that $\beta$ defines the lower boundary on the co-state and is a composite parameter which including terms from profit and from environmental damage: $\beta=\frac{\xi q}{ab}$.
The higher is $\beta$, the more stringent is the environmental regulation and the more stringent is the constraint on the co-state and thus the control.

Hence, as long as the control (being interpreted as the production rate) does not stay on the boundary defined by $\beta$, the environmental damage does not exceed the profit, thus allowing further continued operation. In this way, the notion of an environmentally sustainable strategy seeks a trade-off that combines both profit and environmental concerns, as will be detailed below.

\subsection{Conditions for the optimal solution to be sustainable}
Consider the optimal solution $\lambda^*(t)$ satisfying \eqref{eq:lambda-inf} along with \eqref{eq:u-opt}, and the initial condition $\lambda(0)=\lambda_{eq}<0$. Then, the value of the adjoint variable at the switching time will be equal to $\lambda^*(t_s)=\left(\lambda_{eq}+\frac{1}{\rho_1}\right)e^{\rho_1 t_s} - \frac{1}{\rho_1}<0$. These two values bound the adjoint variable from above and below so that we have $\min\{\lambda_{eq},\lambda^*(t_s)\}\le \lambda^*(t)\le \max\{\lambda_{eq},\lambda^*(t_s)\}<0$.

In the following, we will determine the conditions for the optimal solution to be sustainable, i.e., such that $\lambda^*(t)\in[-1/\beta,0]$ for all $t\ge 0$.

\begin{figure}[tbh]
    \subfloat[$t_s=0.2$\label{fig:a12-1}]{%
      \includegraphics[width=0.33\textwidth]{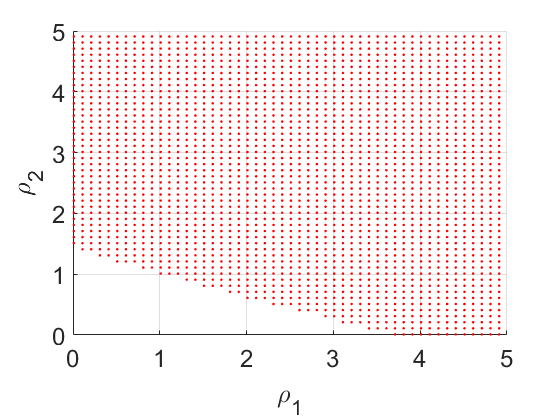}
    }
    \subfloat[$t_s=0.5$\label{fig:a12-2}]{%
      \includegraphics[width=0.33\textwidth]{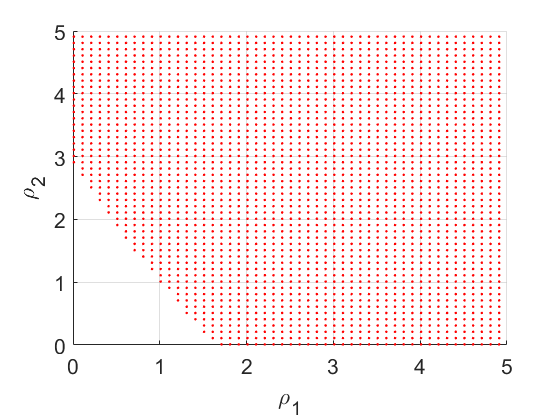}
    }
    \subfloat[$t_s=0.8$\label{fig:a12-3}]{%
      \includegraphics[width=0.33\textwidth]{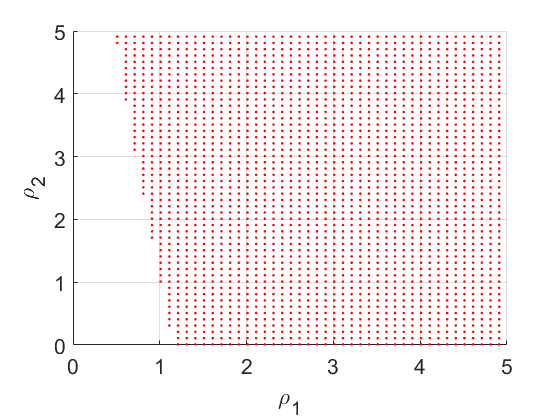}
    }
    \caption{Admissible values of $\rho_1$ and $\rho_2$ for $\beta=1$ and different values of $t_s$.
}
    \label{fig:comparison-inf2}
  \end{figure}

First, we note that the optimal solution $\lambda^*(t)$ follows one of two patterns, depending on the values of $\rho_1$ and $\rho_2$: $\lambda^*(t)$ increases for $t\in [kT,kT+\alpha T)$ and decreases for $t\in [kT+\alpha T,(k+1)T)$ if $\rho_1<\rho_2$ ; alternatively, $\lambda^*(t)$ first decreases and then increases if $\rho_1>\rho_2$. Since $\max\{\lambda_{eq},\lambda^*(t_s)\}<0$, we only need to check the lower bound.
Depending on the values of $\rho_1$ and $\rho_2$, we have to show that 
\begin{subequations}\label{eq:cond-inf}\begin{align}
&\frac 1{\rho_2}+\frac{\rho_2-\rho_1}{\rho_1 \rho_2}\cdot\frac{e^{\rho_ 1 t_s}-1}{e^{\rho_ 1 t_s}-e^{\rho_2 \left(t_s-1\right)}}\le \frac1\beta&\quad\mbox{if}\quad \rho_1<\rho_2\label{eq:cond-inf1}\\
\mbox{or}&\notag\\
&\frac 1{\rho_1}+\frac{\rho_2-\rho_1}{\rho_1 \rho_2}\cdot\frac{e^{\rho_ 1 t_s} \left(e^{\rho_2 \left(t_s-1\right)}-1\right)}{e^{\rho_ 1 t_s}-e^{\rho_2 \left(t_s-1\right)}}
\le \frac1\beta&\quad\mbox{if}\quad \rho_1>\rho_2.\label{eq:cond-inf2}
\end{align}\end{subequations}  
From \eqref{eq:cond-inf} we infer:
\begin{lem}\label{lem:inf}The optimal solution to \eqref{eq:problem} with an infinite number of switches is environmentally sustainable if $\min(\rho_1,\rho_2)\ge \beta$.
\end{lem} 
\begin{proof}
	Suppose that the first phase of the system dynamics is characterized by the smaller absorption coefficient, i.e., $\rho_1<\rho_2$.
	Thus, we consider the first condition \eqref{eq:cond-inf1}.
	It can be readily shown that the value of the expression on the left of \eqref{eq:cond-inf1} changes monotonically from $1/\rho_2$ to $1/\rho_1$ as $t_s$ changes from $0$ to $1$.
	A similar observation can be made for the case \eqref{eq:cond-inf2}, thus yielding the estimate.
\end{proof}
The obtained condition is somewhat conservative, as Figure~\ref{fig:comparison-inf2} illustrates. However, one can see that as $t_s$ grows, the estimate becomes tighter.

The result of Lemma \ref{lem:inf} can be formulated in terms of the original model's parameters.
\begin{cor}The optimal solution to \eqref{eq:problem0} with an infinite number of switches is environmentally sustainable if 
\[\xi q \le ab\,(r+\min(\delta_1,\delta_2)).\]
\end{cor}

Now let us illustrate the concept of environmental sustainability as defined above by taking the values of the impact parameter $\beta$ to be such that the associated control either stays in the interior of $(0,1)$ (i.e., is environmentally sustainable) or otherwise.
\begin{figure}[tbh]
    \subfloat[$\beta=0.8$\label{fig:dyn1}]{%
      \includegraphics[width=0.49\textwidth]{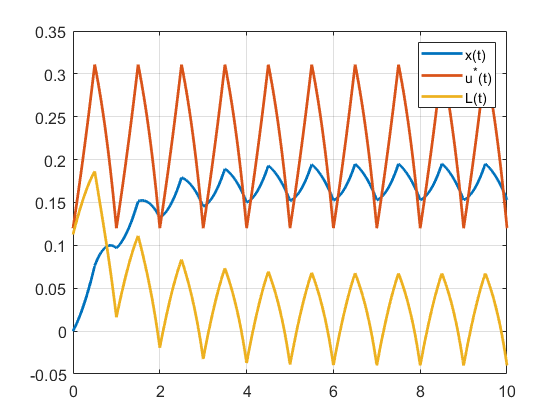}
    }
    \subfloat[$\beta=1$\label{fig:dyn2}]{%
      \includegraphics[width=0.49\textwidth]{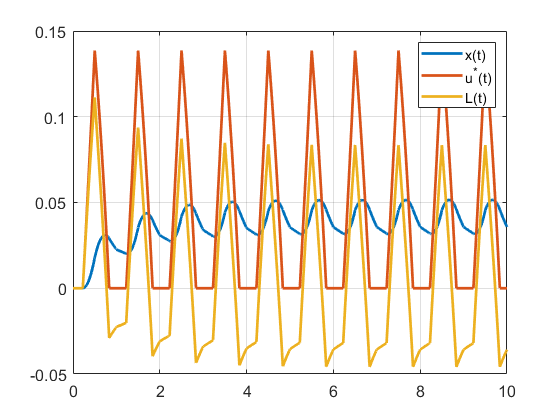}
    }
    \caption{Level of pollution $x(t)$, optimal control $u^*(t)$, and the instantaneous profit $L(t)$ for different values of $\beta$.
Other values of the parameters are set to $x(0)=0$, $\delta_1=0.5$, $\delta_2=1.5$, $r=0.03$, and $t_s=0.5$.}
    \label{fig:dyn}
  \end{figure}

Panel \ref{fig:dyn} shows the optimal control, the evolution of the state, and the instantaneous profit for two types of solutions.
The behavior of the adjoint variable (not shown here) does not depend on $\beta$ and is similar to the plot of the optimal control in Fig.\ \ref{fig:dyn1} (as these two variables are affinely dependent).
The adjoint variable oscillates within the range $[-0.86,-1.1]$, which implies that the optimal control lies within the admissible region for $\beta\le 1/1.1 \approx 0.9$.

Specifically, in Figure \ref{fig:dyn1}, which presents an environmentally sustainable case, it can be seen that the control never turns to zero, so production is ongoing.
The instantaneous profit still can be negative for some periods, but the overall profit remains positive.
The level of pollution, after an initial increase, converges to a stable cycle.

\begin{figure}[thb]
	\centering
	\includegraphics[scale=0.4]{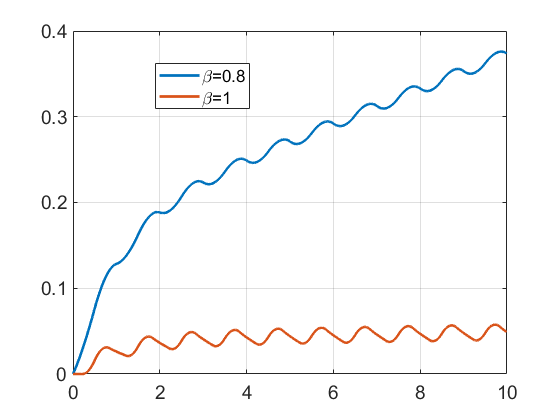}
	\caption{Two payoff functions corresponding to the plots in Fig.~\ref{fig:dyn}.}
	\label{fig:J_2beta}
\end{figure}

Figure \ref{fig:dyn2} illustrates the case of $\beta=1$, where we observe periods of a halt of production ($u(t)=0$).
As a result, there are periods of negative instantaneous profit of  much larger duration than in the case of a sustainable solution.
However, we note that the environment is stabilized at a lower level of pollution than for the case of $\beta=0.8$. 
 This might seem to contradict our previous statement that the environmentally sustainable solution is superior to the saturating one. To clarify this point, we consider Figure~\ref{fig:J_2beta}, where the profit functions are shown for both values of $\beta$. The difference is striking: while the profit function for $\beta=0.8$, i.e., an environmentally sustainable solution is growing, the profit function corresponding to $\beta=1$ oscillates around a relatively low value, thus yielding zero net profit.

To put it differently, an environmentally sustainable solution corresponds to a trade-off between minimizing the level of pollution and guaranteeing the growth of the profit function, whereas the opposite case results in zero on average increase in accumulated profits but lower overall pollution levels. This is in line with the observation that environmental sustainability is closely related to economic sustainability, see \cite{Goodland:95} for a detailed discussion.

However, the described effect appears because of long periods of negative instantaneous payoff. Thus, one may ask whether it is possible to increase the payoff without damaging the environment by stopping the production at the time instants when the instantaneous payoff is negative? Perhaps one can even use this trick to increase the productivity to some extent?
  

\subsection{A myopic solution}
To answer this question, we take a closer look at the profitability of operations.
From \eqref{eq:problem} it immediately follows that the instantaneous profit is nonnegative as long as
\begin{equation}\label{eq:prof}
u\in [1-\sqrt{1-2x},1+\sqrt{1-2x}],\quad \mbox{for}\quad 1-2x\ge 0.
\end{equation}
We call the optimal solution respecting this constraint set on control {\emph myopic} as the decision maker chooses to deviate from the optimal solution in an attempt to avoid short-term losses. This interpretation somewhat deviates from the use of the word ``myopic'' in differential game theory. However, we believe that it still retains the main meaning of preferring short-term advantages over the long-term ones. Since this paper is not concerned with differential game theory, there should be no room for confusion in this terminology. Note that the described solution can be alternatively termed {\em liquidity constrained}, but the latter term seems to be overly specialized.

\begin{figure}[thb]
	\centering
	\includegraphics[scale=0.4]{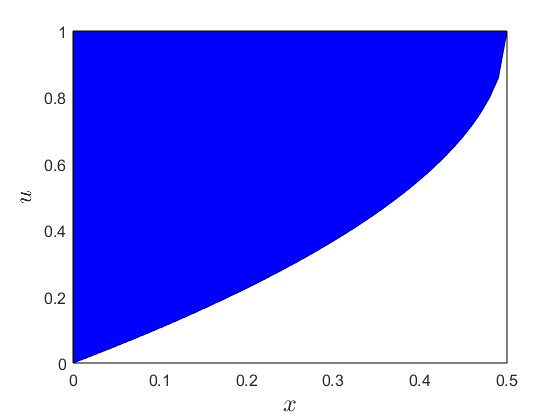}
	\caption{The region of $1-\sqrt{1-2x}\leq u$}
	\label{uxreg}
\end{figure}

Note that if $1-2x<0$, the control takes its maximal value, which does not suffice to keep the profit positive: it becomes negative for some period.
Furthermore, we have $1+\sqrt{1-2x}\geq 1$ so only the \emph{lower} boundary for the control is essential, and it is stricter  than the previously defined boundary $u\ge 0$.
Figure \ref{uxreg} gives the region of $u(x)$ granting a positive per-period profit stream.

Condition \eqref{eq:prof} describes the \emph{liquidity constraint:} if a firm is allowed to borrow from previous, resp., future periods, it still can cope with periods of negative profit if the total profit stream remains positive, and then the satisfaction of the sustainability condition is enough for a seamless operation.
However, if there are constraints and the firm cannot carry over funds from period to period, the condition \eqref{eq:prof} has to be respected at all times.

Now, this is the question for policy regulation: recall that the initial problem \eqref{eq:J0} includes the ecotax parameter $q$.
The size of this would define how stringent is the environmental constraint, and thus the relative weight of the pollution stock $x$ in the net profit.
It could be suggested that a more flexible policy, with $q$ varying with the same seasonality as $\delta$, would smooth over seasonal profit variations and this could be helpful in situations with binding liquidity constraints. However, a more detailed treatment of this regulation problem is left for future research.

We rather focus on the comparison of such a \emph{myopic} solution with the sustainable one.
To this end, we first formulate the strategy associated with constraint \eqref{eq:prof} and compare it with the previous one over short and long-time horizons.

\begin{figure}[tbh]
    \subfloat[$\beta=0.8$\label{fig:dyn1x}]{%
      \includegraphics[width=0.49\textwidth]{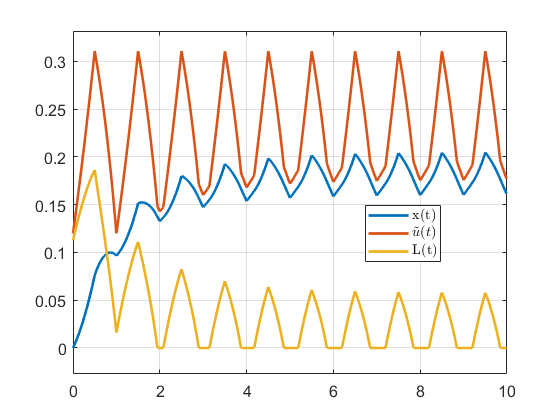}
    }
    \subfloat[$\beta=1$\label{fig:dyn2x}]{%
      \includegraphics[width=0.49\textwidth]{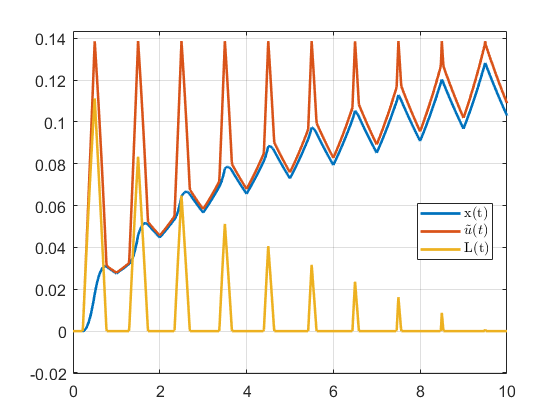}
    }
    \caption{Level of pollution $x(t)$, myopic control $\tilde{u}(t)$, and the instantaneous profit $L(t)$ for different values of $\beta$ for the case when the control is determined according to \eqref{eq:u-opt-constr}.
Other values of the parameters are set to $x(0)=0$, $\delta_1=0.5$, $\delta_2=1.5$, $r=0.03$, and $t_s=0.5$.}
    \label{fig:dynx}
  \end{figure}

 We thus modify \eqref{eq:u-opt} as follows:
\begin{equation}\label{eq:u-opt-constr}
\tilde{u}(t)=\begin{cases}
1-\sqrt{1-2x},& -\beta\lambda(t)\ge \sqrt{1-2x}\; \wedge\; x\le \frac12\\
1,&\lambda(t)>0\; \vee \; x>\frac12\\
1+\beta\lambda(t),& \mbox{otherwise}.
\end{cases}
\end{equation}
The resulting optimal control problem is well-defined, as the dynamics of $\lambda(t)$ is decoupled from both the state $x(t)$ and the  control $u(t)$.
The resulting plots are presented in Panel \ref{fig:dynx}.
  
\begin{figure}[tbh]
    \subfloat[Myopic solution\label{fig:J2-long}]{%
      \includegraphics[width=0.49\textwidth]{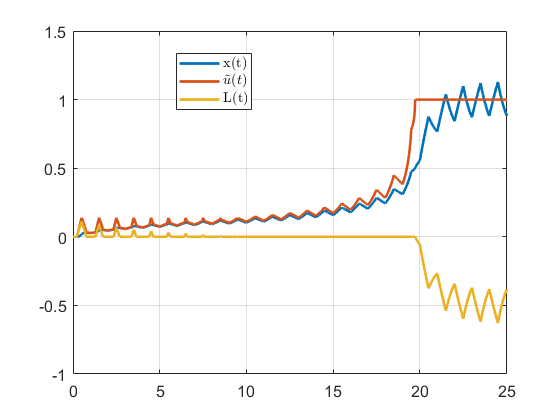}
    }
    \subfloat[Comparison of profit functions\label{fig:J2-cost}]{%
      \includegraphics[width=0.49\textwidth]{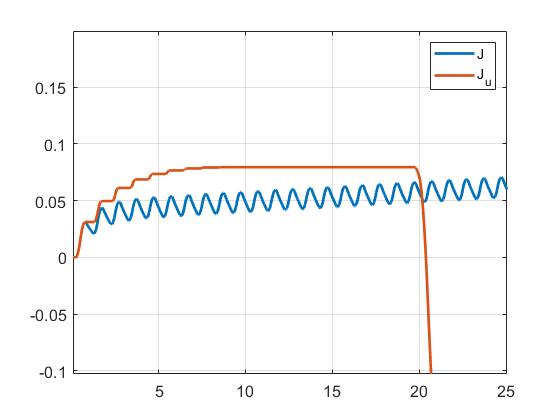}
    }
    \caption{In this panel, we illustrate the myopic solution \eqref{eq:u-opt-constr} (left) and compare the two payoff functions for the optimal control \eqref{eq:u-opt} (labelled $J$) and the control \eqref{eq:u-opt-constr} (labelled $J_{u}$) (right) with $\beta=1$.}
    \label{fig:J2}
  \end{figure}

  We can observe that while in Fig.\ \ref{fig:dyn1x} the application of a new control leads to a (slightly) increased level of pollution, the situation presented in Fig.\ \ref{fig:dyn2x} differs drastically from its counterpart in Fig.\ \ref{fig:dyn2}.
Namely, in an attempt to equilibrate the profit and the expenses, the decision maker has to increase the level of production, which in turn leads to an increase of the ecotax and so on.
Following this strategy, we eventually arrive at the situation where one has to increase the rate of production (resp., the rate of pollution) only to achieve zero net profit.
Clearly, this strategy is unsustainable in both aspects: the level of pollution increases, while the stream of profit decreases to zero.
Even worse, as the level of pollution grows beyond $1/2$, the control cannot compensate for the related fines any longer and the payoff starts decreasing.
This situation is illustrated in Fig.\ \ref{fig:J2}, where we show the long-term effect of choosing a myopic solution.
It is interesting to note that in the short term, the solution \eqref{eq:u-opt-constr} does indeed perform better than \eqref{eq:u-opt}.
This is caused by the fact that the solution \eqref{eq:u-opt} is optimal on an infinite horizon, but fails to be optimal on a finite (and short-term) horizon.

\begin{figure}[tbh]
    \subfloat[ $t\in{[0,10]}$ \label{fig:J1a}]{%
      \includegraphics[width=0.49\textwidth]{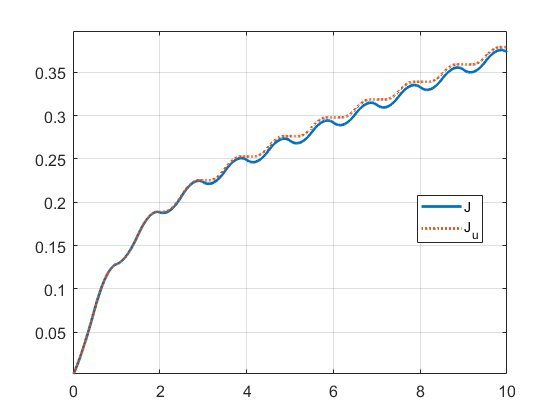}
    }
    \subfloat[$t\in{[30,40]}$\label{fig:J1b}]{%
      \includegraphics[width=0.49\textwidth]{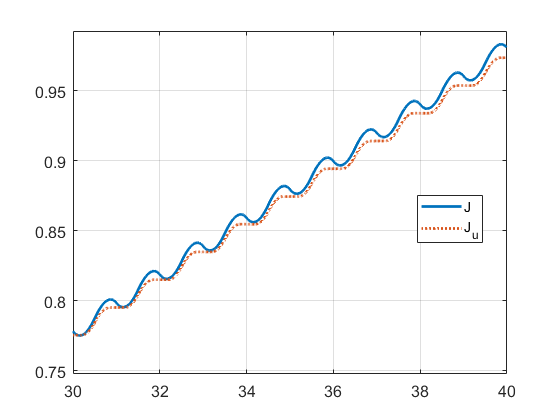}
    }
    \caption{Comparison of two payoff functions for $\beta=0.8$ over two different intervals of time: $[0,10]$ (left), and $[30,40]$ (right).}
    \label{fig:J1}
  \end{figure}

The previous observation is supported by the results obtained for the case $\beta=0.8$, as shown in Fig.\ \ref{fig:J1}.
At first, it may seem that solution \eqref{eq:u-opt-constr} yields a better result, but as time goes on, the solution \eqref{eq:u-opt} overtakes it.

We conclude this comparison with the following observation.
It might seem that one can still gain by using the myopic solution for a short time and then switching to the optimal one.
For instance, Fig.~\ref{fig:J2} may suggest one stay with $\tilde{u}(t)$ until $t=5$ and then switch to $u^*(t)$.
However, we can easily see that at $t=5$, the value of the state variable is already much higher than the optimal one.
So, switching to the optimal solution would imply that the control has to stay equal to $0$ for some time to let nature reduce the accumulated pollution.
Obviously, during this time there is no production and the profit decreases, thus neutralizing the achieved improvement.

\section{Conclusions}\label{sec:5}
In this paper, we studied the hybrid optimal control problem of pollution control with linear dependence of ecotax on the state.

It has been shown that under seasonal fluctuations of $\delta$, the regenerative capacity of the environment, there is a unique optimal solution that has the form of a hybrid limit cycle. This is, to the best of our knowledge,  the first such result in the literature. We thus further explored environmental and policy implications of such an optimal policy.

We next determine the conditions on the coefficients of the model, guaranteeing that the optimal solution lies within the admissible bounds (the environmentally sustainable solution).
The latter requirement is shown to provide an optimal balance between the profit due to the production and the expenses due to the ecological tax. Again, to our knowledge, this is the first attempt to define an environmentally sustainable solution in rigorous mathematical terms of the underlying model's dynamics.

We illustrate this thesis by considering a myopic solution as the one with additional constraint on the nonnegativity of instantaneous profit and compare such a solution with the optimal one.
Interestingly enough, in the short term this solution provides a better payoff, but at the cost of environmental deterioration reminding of myopic behavior of a kind.
However, in the long term, the sustainable solution provides higher payoff due to the more conservative use of environmental resources (production is lower and stays limited).
At the same time, the sustainable solution allows avoiding periodic halts in production.
We discuss the effect of the system parameters on the type of the optimal solution and indicate ways to modify the parameters of the system to ensure that the obtained optimal strategy is environmentally sustainable.

The class of problems studied here can be extended to include many particular applications that undergo seasonal variations.
These range from renewable energy production (where the day/night cycle affects the efficiency of energy generation) to pollution management and seasonal market demand fluctuations.
Another possible extension consists in considering this problem within the class of multi-agent differential games.

Finally, we note that the environmental sustainability condition presented in this paper reflects only one aspect of a much more complicated process. In particular, we assumed that while the regenerative ability of the environment is changing with time, the overall capacity of the reservoir is infinite. However, it might be possible that at a certain level of pollution the reservoir switches to a different state, which can be characterized by entirely different dynamics. One possible example is the shallow lake model described in detail in \cite{GU:21}, which switches from an oligotrophic to a eutrophic state depending on the level of pollutant.

\section*{Appendix}
\begin{proof}{\bf Theorem~\ref{thm1}.} Note that the solution satisfying $\lambda(0)=\lambda_{eq}$ is the only bounded solution to \eqref{eq:lambda-inf} (see \cite[Lemma 3]{GBG:21OL}). Along with the boundedness of the state $x(t)$, this guarantees the fulfillment of the transversality condition $\lim_{t\to \infty} e^{-r t} \lambda(t)(x(t)-x^*(t)) = 0$ for any admissible solution $x(t)$. The latter, along with concavity of the Hamiltonian w.r.t. the state $x$, guarantees that the pair $(x^*(t),u^*(t))$ is an optimal solution to \eqref{eq:problem}, \eqref{eq.delta} (cf. \cite[Theorem 13]{Seierstad/Sydsaeter:1987}, \cite[Theorem 13.1]{AK:07}). Now, the uniqueness of the obtained optimal solution follows from the concavity of the Hamiltonian function w.r.t. the state $x$ and its strict concavity w.r.t. the control $u$. 
\end{proof}

\begin{proof}{\bf Theorem~\ref{thm2}.} Consider the differential equation 
\begin{equation}\label{eq:sysA}\dot{x}=\beta u^*(t) - \delta(t)x(t),\quad x(0)=x_0,\tag{A1}\end{equation}
with $u^*(t)$ and $\delta(t)$ are defined by \eqref{eq:u-opt} and \eqref{eq.delta}. Let us define the function $f(t)$ as
\[f(t)=e^{-\int_0^t \delta(s)ds}\int\limits_0^t \beta u^*(\tau)e^{\int_0^\tau \delta(s)ds}d\tau.\]
The function $f(t)$ satisfies
\[\dot{f}(t)=\beta u^*(t)-\delta(t)f(t),\quad f(0)=0.\]
Let, furthermore, the function $\phi(t,x_0)$ be defined as
\[\phi(t,x_0)=x_0 e^{-\int_0^t \delta(s)ds} +f(t).\]
It holds that $\phi(0,x_0)=x_0$ and
\[\begin{aligned}\frac{\partial}{\partial t}\phi(t,x_0)=&\,-\delta(t)x_0 e^{-\int_0^t \delta(s)ds} + \beta u^*(t)-\delta(t)f(t)\\
 =&\, \beta u^*(t) - \delta(t)\phi(t,x_0). \end{aligned}\]
The latter implies that the function $\phi(t,x_0)$ is the solution of the initial value problem \eqref{eq:sysA}. Since the function $f(t)$ does not depend on the initial condition $x_0$, we have
\[\phi(t,x'_0)-\phi(t,x''_0)=(x'_0-x''_0) e^{-\int_0^t \delta(s)ds}.\]
Noting that $\delta(t)\ge \delta_{\min}=\min(\delta_1,\delta_2)>0$, we obtain the following bound for any two initial values $x'_0$ and $x''_0$:
\begin{equation}\label{eq:ineq-phi}|\phi(t,x'_0)-\phi(t,x''_0)|=|x'_0-x''_0| e^{-\int_0^t \delta(s)ds}\le |x'_0-x''_0|e^{-\delta_{\min}t}.\tag{A2}\end{equation}
Now choose an arbitrary positive constant $\bar{x}>\frac{\beta}{\delta_{\min}}$ and define $X=[0,\bar{x}]$. Then, by Lemma~\ref{lem1}, $\phi(T,x)\in X$ for any $x\in X$. 

Define the mapping $S:X\to X$ as $S(x)=\phi(T,x)$. By Brouwer's fixed-point theorem, there exists a solution $x_{eq}$ satisfying $x_{eq}=\phi(T,x_{eq})$. Furthermore, by \eqref{eq:ineq-phi},
\[|S(x')-S(x'')|\le |x'-x''|e^{-\delta_{\min}T}<|x'-x''|,\]
which implies that $S$ is a contraction mapping and hence, $x_{eq}$ is unique.

Thus, the solution to \eqref{eq:sysA} satisfying $x(0)=x_{eq}$ constitutes a unique hybrid limit cycle in the state variable, denote it $x_h(t)$. Moreover, by the above arguments any solution with $x(0)\in X$ converges to $x_h(t)$ as $t$ goes to $\infty$. Finally, as the upper bound of $X$ can be set to be arbitrarily large, any solution with $x(0)>0$ converges to $x_h(t)$ as well.
\end{proof}

\section*{Funding}

The work of A.~Bondarev was supported by the Key Program Special Fund  of Xi'an Jiaotong-Liverpool University (grant no. KSF-E-63).
The work of D.~Gromov was supported by the RFBR and DFG, project number 21-51-12007.

\section*{Data availability} All data generated or analyzed during this study are included in this published article.


\bibliographystyle{cas-model2-names}
\bibliography{GSB}

\begin{thebibliography}{39}
\expandafter\ifx\csname natexlab\endcsname\relax\def\natexlab#1{#1}\fi
\providecommand{\url}[1]{\texttt{#1}}
\providecommand{\href}[2]{#2}
\providecommand{\path}[1]{#1}
\providecommand{\DOIprefix}{doi:}
\providecommand{\ArXivprefix}{arXiv:}
\providecommand{\URLprefix}{URL: }
\providecommand{\Pubmedprefix}{pmid:}
\providecommand{\doi}[1]{\href{http://dx.doi.org/#1}{\path{#1}}}
\providecommand{\Pubmed}[1]{\href{pmid:#1}{\path{#1}}}
\providecommand{\bibinfo}[2]{#2}
\ifx\xfnm\relax \def\xfnm[#1]{\unskip,\space#1}\fi
\bibitem[{Arguedas et~al.(2017)Arguedas, Cabo and
  Mart{\'\i}n-Herr{\'a}n}]{Arguedas:17}
\bibinfo{author}{Arguedas, C.}, \bibinfo{author}{Cabo, F.},
  \bibinfo{author}{Mart{\'\i}n-Herr{\'a}n, G.}, \bibinfo{year}{2017}.
\newblock \bibinfo{title}{Optimal pollution standards and non-compliance in a
  dynamic framework}.
\newblock \bibinfo{journal}{Environmental and Resource Economics}
  \bibinfo{volume}{68}, \bibinfo{pages}{537--567}.
\bibitem[{Arguedas et~al.(2020)Arguedas, Cabo and
  Mart{\'\i}n-Herr{\'a}n}]{Arguedas:20}
\bibinfo{author}{Arguedas, C.}, \bibinfo{author}{Cabo, F.},
  \bibinfo{author}{Mart{\'\i}n-Herr{\'a}n, G.}, \bibinfo{year}{2020}.
\newblock \bibinfo{title}{Enforcing regulatory standards in stock pollution
  problems}.
\newblock \bibinfo{journal}{Journal of Environmental Economics and Management}
  \bibinfo{volume}{100}, \bibinfo{pages}{102297}.
\bibitem[{Aseev and Kryazhimskii(2007)}]{AK:07}
\bibinfo{author}{Aseev, S.M.}, \bibinfo{author}{Kryazhimskii, A.V.},
  \bibinfo{year}{2007}.
\newblock \bibinfo{title}{The pontryagin maximum principle and optimal economic
  growth problems}.
\newblock \bibinfo{journal}{Proceedings of the Steklov institute of
  mathematics} \bibinfo{volume}{257}, \bibinfo{pages}{1--255}.
\bibitem[{Belyakov et~al.(2015)Belyakov, Davydov and Veliov}]{BDV:2015}
\bibinfo{author}{Belyakov, A.O.}, \bibinfo{author}{Davydov, A.A.},
  \bibinfo{author}{Veliov, V.M.}, \bibinfo{year}{2015}.
\newblock \bibinfo{title}{Optimal cyclic exploitation of renewable resources}.
\newblock \bibinfo{journal}{Journal of Dynamical and Control Systems}
  \bibinfo{volume}{21}, \bibinfo{pages}{475--494}.
\bibitem[{Belyakov and Veliov(2014)}]{Belyakov/Veliov:2014}
\bibinfo{author}{Belyakov, A.O.}, \bibinfo{author}{Veliov, V.M.},
  \bibinfo{year}{2014}.
\newblock \bibinfo{title}{Constant versus periodic fishing: Age structured
  optimal control approach}.
\newblock \bibinfo{journal}{Mathematical Modelling of Natural Phenomena}
  \bibinfo{volume}{9}, \bibinfo{pages}{20--37}.
\bibitem[{Bondarev and Gromov(2021)}]{BG19}
\bibinfo{author}{Bondarev, A.}, \bibinfo{author}{Gromov, D.},
  \bibinfo{year}{2021}.
\newblock \bibinfo{title}{On the structure and regularity of optimal solutions
  in a differential game with regime switching and spillovers}, in:
  \bibinfo{booktitle}{Dynamic economic models with regime switches}.
  \bibinfo{publisher}{Springer International Publishing},
  \bibinfo{address}{Cham}, pp. \bibinfo{pages}{187--207}.
\bibitem[{Breton et~al.(2005)Breton, Zaccour and Zahaf}]{BZZ:05}
\bibinfo{author}{Breton, M.}, \bibinfo{author}{Zaccour, G.},
  \bibinfo{author}{Zahaf, M.}, \bibinfo{year}{2005}.
\newblock \bibinfo{title}{A differential game of joint implementation of
  environmental projects}.
\newblock \bibinfo{journal}{Automatica} \bibinfo{volume}{41},
  \bibinfo{pages}{1737--1749}.
\bibitem[{Caines et~al.(2007)Caines, Egerstedt, Malham\'e and
  Schoellig}]{CEMS:07}
\bibinfo{author}{Caines, P.E.}, \bibinfo{author}{Egerstedt, M.},
  \bibinfo{author}{Malham\'e, R.}, \bibinfo{author}{Schoellig, A.},
  \bibinfo{year}{2007}.
\newblock \bibinfo{title}{A hybrid {B}ellman equation for bimodal systems}, in:
  \bibinfo{editor}{Bemporad, A.}, \bibinfo{editor}{Bicchi, A.},
  \bibinfo{editor}{Butazzo, G.} (Eds.), \bibinfo{booktitle}{HSCC 2007}.
  \bibinfo{publisher}{Springer}. LNCS 4416, pp. \bibinfo{pages}{656--659}.
\bibitem[{Chev{\'e}(2000)}]{Cheve:00}
\bibinfo{author}{Chev{\'e}, M.}, \bibinfo{year}{2000}.
\newblock \bibinfo{title}{Irreversibility of pollution accumulation}.
\newblock \bibinfo{journal}{Environmental and Resource Economics}
  \bibinfo{volume}{16}, \bibinfo{pages}{93--104}.
\newblock \DOIprefix\doi{10.1023/A:1008367226371}.
\bibitem[{De~Zeeuw and Zemel(2012)}]{deZeeuw:12}
\bibinfo{author}{De~Zeeuw, A.}, \bibinfo{author}{Zemel, A.},
  \bibinfo{year}{2012}.
\newblock \bibinfo{title}{Regime shifts and uncertainty in pollution control}.
\newblock \bibinfo{journal}{Journal of Economic Dynamics and Control}
  \bibinfo{volume}{36}, \bibinfo{pages}{939--950}.
\bibitem[{Dockner et~al.(2000)Dockner, Jorgensen, Long and Sorger}]{Dockner}
\bibinfo{author}{Dockner, E.}, \bibinfo{author}{Jorgensen, S.},
  \bibinfo{author}{Long, N.}, \bibinfo{author}{Sorger, G.},
  \bibinfo{year}{2000}.
\newblock \bibinfo{title}{Differential Games in Economics and Management
  Sciences}.
\newblock \bibinfo{publisher}{Cambridge University Press, Cambridge}.
\bibitem[{El~Ouardighi et~al.(2014)El~Ouardighi, Benchekroun and
  Grass}]{OBG:14}
\bibinfo{author}{El~Ouardighi, F.}, \bibinfo{author}{Benchekroun, H.},
  \bibinfo{author}{Grass, D.}, \bibinfo{year}{2014}.
\newblock \bibinfo{title}{Controlling pollution and environmental absorption
  capacity}.
\newblock \bibinfo{journal}{Annals of Operations Research}
  \bibinfo{volume}{220}, \bibinfo{pages}{111--133}.
\newblock \DOIprefix\doi{10.1007/s10479-011-0982-4}.
\bibitem[{Goodland(1995)}]{Goodland:95}
\bibinfo{author}{Goodland, R.}, \bibinfo{year}{1995}.
\newblock \bibinfo{title}{The concept of environmental sustainability}.
\newblock \bibinfo{journal}{Annual review of ecology and systematics} ,
  \bibinfo{pages}{1--24}.
\bibitem[{Gromov et~al.(2021)Gromov, Bondarev and Gromova}]{GBG:21OL}
\bibinfo{author}{Gromov, D.}, \bibinfo{author}{Bondarev, A.},
  \bibinfo{author}{Gromova, E.}, \bibinfo{year}{2021}.
\newblock \bibinfo{title}{On periodic solution to control problem with
  time-driven switching}.
\newblock \bibinfo{journal}{Optimization Letters}
  \DOIprefix\doi{10.1007/s11590-021-01749-6}. \bibinfo{note}{on-line first.}
\bibitem[{Gromov and Gromova(2017)}]{Gromov2017}
\bibinfo{author}{Gromov, D.}, \bibinfo{author}{Gromova, E.},
  \bibinfo{year}{2017}.
\newblock \bibinfo{title}{On a class of hybrid differential games}.
\newblock \bibinfo{journal}{Dynamic Games and Applications}
  \bibinfo{volume}{7}, \bibinfo{pages}{266--288}.
\newblock \DOIprefix\doi{10.1007/s13235-016-0185-3}.
\bibitem[{Gromov and Upmann(2021)}]{GU:21}
\bibinfo{author}{Gromov, D.}, \bibinfo{author}{Upmann, T.},
  \bibinfo{year}{2021}.
\newblock \bibinfo{title}{Dynamics and economics of shallow lakes: A survey}.
\newblock \bibinfo{journal}{Sustainability} \bibinfo{volume}{13}.
\newblock \DOIprefix\doi{10.3390/su132413763}.
\bibitem[{Hoekstra and van~den Bergh(2005)}]{Hoekstra/Bergh:2005}
\bibinfo{author}{Hoekstra, J.}, \bibinfo{author}{van~den Bergh, J.C.J.M.},
  \bibinfo{year}{2005}.
\newblock \bibinfo{title}{Harvesting and conservation in a predator-prey
  system}.
\newblock \bibinfo{journal}{Journal of Economic Dynamics and Control}
  \bibinfo{volume}{29}, \bibinfo{pages}{1097--1120}.
\bibitem[{J{\o}rgensen et~al.(2010)J{\o}rgensen, Mart{\'\i}n-Herr{\'a}n and
  Zaccour}]{Jorg:10}
\bibinfo{author}{J{\o}rgensen, S.}, \bibinfo{author}{Mart{\'\i}n-Herr{\'a}n,
  G.}, \bibinfo{author}{Zaccour, G.}, \bibinfo{year}{2010}.
\newblock \bibinfo{title}{Dynamic games in the economics and management of
  pollution}.
\newblock \bibinfo{journal}{Environmental Modeling \& Assessment}
  \bibinfo{volume}{15}, \bibinfo{pages}{433--467}.
\newblock \DOIprefix\doi{10.1007/s10666-010-9221-7}.
\bibitem[{Jouvet et~al.(2005)Jouvet, Michel and Rotillon}]{Jouvet:05}
\bibinfo{author}{Jouvet, P.A.}, \bibinfo{author}{Michel, P.},
  \bibinfo{author}{Rotillon, G.}, \bibinfo{year}{2005}.
\newblock \bibinfo{title}{Optimal growth with pollution: how to use pollution
  permits?}
\newblock \bibinfo{journal}{Journal of Economic Dynamics and Control}
  \bibinfo{volume}{29}, \bibinfo{pages}{1597--1609}.
\bibitem[{Klamerus-Iwan et~al.(2018)Klamerus-Iwan, B{\l}o{\'n}ska, Lasota,
  Walig{\'o}rski and Kalandyk}]{Klamerus:18}
\bibinfo{author}{Klamerus-Iwan, A.}, \bibinfo{author}{B{\l}o{\'n}ska, E.},
  \bibinfo{author}{Lasota, J.}, \bibinfo{author}{Walig{\'o}rski, P.},
  \bibinfo{author}{Kalandyk, A.}, \bibinfo{year}{2018}.
\newblock \bibinfo{title}{Seasonal variability of leaf water capacity and
  wettability under the influence of pollution in different city zones}.
\newblock \bibinfo{journal}{Atmospheric Pollution Research}
  \bibinfo{volume}{9}, \bibinfo{pages}{455--463}.
\bibitem[{Liski et~al.(2001)Liski, Kort and Novak}]{LKN:2001}
\bibinfo{author}{Liski, M.}, \bibinfo{author}{Kort, P.M.},
  \bibinfo{author}{Novak, A.}, \bibinfo{year}{2001}.
\newblock \bibinfo{title}{Increasing returns and cycles in fishing}.
\newblock \bibinfo{journal}{Resource and Energy Economics}
  \bibinfo{volume}{23}, \bibinfo{pages}{241--258}.
\bibitem[{Liu et~al.(2018)Liu, Yu, Xie, Von~Gadow and Peng}]{Liu:18}
\bibinfo{author}{Liu, W.}, \bibinfo{author}{Yu, Z.}, \bibinfo{author}{Xie, X.},
  \bibinfo{author}{Von~Gadow, K.}, \bibinfo{author}{Peng, C.},
  \bibinfo{year}{2018}.
\newblock \bibinfo{title}{A critical analysis of the carbon neutrality
  assumption in life cycle assessment of forest bioenergy systems}.
\newblock \bibinfo{journal}{Environmental Reviews} \bibinfo{volume}{26},
  \bibinfo{pages}{93--101}.
\newblock \DOIprefix\doi{10.1139/er-2017-0060}.
\bibitem[{Luqman et~al.(2018)Luqman, Peng, Huang, Bibi and Najid}]{Luqman:18}
\bibinfo{author}{Luqman, M.}, \bibinfo{author}{Peng, S.},
  \bibinfo{author}{Huang, S.}, \bibinfo{author}{Bibi, A.},
  \bibinfo{author}{Najid, A.}, \bibinfo{year}{2018}.
\newblock \bibinfo{title}{Cost allocation for the problem of pollution
  reduction: a dynamic cooperative game approach}.
\newblock \bibinfo{journal}{Economic research-Ekonomska istra{\v{z}}ivanja}
  \bibinfo{volume}{31}, \bibinfo{pages}{1717--1736}.
\newblock \DOIprefix\doi{10.1080/1331677X.2018.1515642}.
\bibitem[{M{\"a}ler et~al.(2003)M{\"a}ler, Xepapadeas and De~Zeeuw}]{MXZ:03}
\bibinfo{author}{M{\"a}ler, K.G.}, \bibinfo{author}{Xepapadeas, A.},
  \bibinfo{author}{De~Zeeuw, A.}, \bibinfo{year}{2003}.
\newblock \bibinfo{title}{The economics of shallow lakes}.
\newblock \bibinfo{journal}{Environmental and resource Economics}
  \bibinfo{volume}{26}, \bibinfo{pages}{603--624}.
\newblock \DOIprefix\doi{10.1023/B:EARE.0000007351.99227.42}.
\bibitem[{Moberg et~al.(2019)Moberg, Pinsky and Fenichel}]{MPF:2019}
\bibinfo{author}{Moberg, E.A.}, \bibinfo{author}{Pinsky, M.L.},
  \bibinfo{author}{Fenichel, E.P.}, \bibinfo{year}{2019}.
\newblock \bibinfo{title}{Capital investment for optimal exploitation of
  renewable resource stocks in the age of global change}.
\newblock \bibinfo{journal}{Ecological Economics} \bibinfo{volume}{165},
  \bibinfo{pages}{106335}.
\bibitem[{Nieuwenhuijsen et~al.(2007)Nieuwenhuijsen, Gomez-Perales and
  Colvile}]{Nieuwenhuijsen:07}
\bibinfo{author}{Nieuwenhuijsen, M.}, \bibinfo{author}{Gomez-Perales, J.},
  \bibinfo{author}{Colvile, R.}, \bibinfo{year}{2007}.
\newblock \bibinfo{title}{Levels of particulate air pollution, its elemental
  composition, determinants and health effects in metro systems}.
\newblock \bibinfo{journal}{Atmospheric Environment} \bibinfo{volume}{41},
  \bibinfo{pages}{7995--8006}.
\newblock \DOIprefix\doi{10.1016/j.atmosenv.2007.08.002}.
\bibitem[{Nkuiya and Costello(2016)}]{Nkuiya:16}
\bibinfo{author}{Nkuiya, B.}, \bibinfo{author}{Costello, C.},
  \bibinfo{year}{2016}.
\newblock \bibinfo{title}{Pollution control under a possible future shift in
  environmental preferences}.
\newblock \bibinfo{journal}{Journal of Economic Behavior \& Organization}
  \bibinfo{volume}{132}, \bibinfo{pages}{193--205}.
\bibitem[{OECD(2011)}]{oecd:11}
\bibinfo{author}{OECD}, \bibinfo{year}{2011}.
\newblock \bibinfo{title}{Environmental Taxation. {A} Guide for Policy Makers}.
\newblock \bibinfo{type}{Technical Report}. Organisation for Economic
  Co-operation and Development.
\newblock \URLprefix
  \url{https://www.oecd.org/env/tools-evaluation/48164926.pdf}.
\bibitem[{Pichika and Zawka(2018)}]{Pichika/Zawka:2018}
\bibinfo{author}{Pichika, S.D.N.}, \bibinfo{author}{Zawka, S.D.},
  \bibinfo{year}{2018}.
\newblock \bibinfo{title}{Renewable resource management in a seasonally
  fluctuating environment with restricted harvesting effort}.
\newblock \bibinfo{journal}{Mathematical Biosciences} \bibinfo{volume}{301},
  \bibinfo{pages}{1--9}.
\bibitem[{Pichika and Zawka(2019)}]{Pichika/Zawka:2019}
\bibinfo{author}{Pichika, S.D.N.}, \bibinfo{author}{Zawka, S.D.},
  \bibinfo{year}{2019}.
\newblock \bibinfo{title}{Optimal harvesting of a renewable resource in a
  polluted environment: An allocation problem of the sole owner}.
\newblock \bibinfo{journal}{Natural Resource Modeling} \bibinfo{volume}{32}.
\bibitem[{Reddy et~al.(2020)Reddy, Schumacher and Engwerda}]{RG:19}
\bibinfo{author}{Reddy, P.V.}, \bibinfo{author}{Schumacher, J.M.},
  \bibinfo{author}{Engwerda, J.}, \bibinfo{year}{2020}.
\newblock \bibinfo{title}{Analysis of optimal control problems for hybrid
  systems with one state variable}.
\newblock \bibinfo{journal}{SIAM Journal on Control and Optimisation}
  \bibinfo{volume}{58}, \bibinfo{pages}{3262--3292}.
\newblock \DOIprefix\doi{https://doi.org/10.1137/19M1272779}.
\bibitem[{Savkin and Matveev(2000)}]{matveev2000}
\bibinfo{author}{Savkin, A.}, \bibinfo{author}{Matveev, A.},
  \bibinfo{year}{2000}.
\newblock \bibinfo{title}{Cyclic linear differential automata: A simple class
  of hybrid dynamical systems}.
\newblock \bibinfo{journal}{Automatica} \bibinfo{volume}{36},
  \bibinfo{pages}{727--734}.
\newblock \DOIprefix\doi{10.1016/S0005-1098(99)00199-5}.
\bibitem[{Savkin and Matveev(1999)}]{Savkin:99}
\bibinfo{author}{Savkin, A.}, \bibinfo{author}{Matveev, A.S.},
  \bibinfo{year}{1999}.
\newblock \bibinfo{title}{Qualitative analysis of differential automata:
  {E}xistence and stability of limit cycles}, in: \bibinfo{booktitle}{Proc. of
  the Information, Decision and Control Symposium},
  \bibinfo{organization}{IEEE}. pp. \bibinfo{pages}{265--270}.
\bibitem[{Schoellig et~al.(2007)Schoellig, Caines, Egerstedt and
  Malham{\'e}}]{Schollig:07}
\bibinfo{author}{Schoellig, A.}, \bibinfo{author}{Caines, P.E.},
  \bibinfo{author}{Egerstedt, M.}, \bibinfo{author}{Malham{\'e}, R.},
  \bibinfo{year}{2007}.
\newblock \bibinfo{title}{A hybrid bellman equation for systems with regional
  dynamics}, in: \bibinfo{booktitle}{2007 46th IEEE Conference on Decision and
  Control}, \bibinfo{organization}{IEEE}. pp. \bibinfo{pages}{3393--3398}.
\bibitem[{Seidl(2019)}]{Seidl:19}
\bibinfo{author}{Seidl, A.}, \bibinfo{year}{2019}.
\newblock \bibinfo{title}{Zeno points in optimal control models with endogenous
  regime switching}.
\newblock \bibinfo{journal}{Journal of Economic Dynamics and Control}
  \bibinfo{volume}{100}, \bibinfo{pages}{353--368}.
\bibitem[{Seierstad and Syds{\ae}ter(1987)}]{Seierstad/Sydsaeter:1987}
\bibinfo{author}{Seierstad, A.}, \bibinfo{author}{Syds{\ae}ter, K.},
  \bibinfo{year}{1987}.
\newblock \bibinfo{title}{{Optimal Control Theory with Economic Applications}}.
  volume~\bibinfo{volume}{24} of \textit{\bibinfo{series}{Advanced Textbooks in
  Economics}}.
\newblock \bibinfo{publisher}{North-Holland}, \bibinfo{address}{Amsterdam}.
\bibitem[{Shortle and Horan(2001)}]{Shortle:01}
\bibinfo{author}{Shortle, J.S.}, \bibinfo{author}{Horan, R.D.},
  \bibinfo{year}{2001}.
\newblock \bibinfo{title}{The economics of nonpoint pollution control}.
\newblock \bibinfo{journal}{Journal of economic surveys} \bibinfo{volume}{15},
  \bibinfo{pages}{255--289}.
\bibitem[{Wang et~al.(2013)Wang, Shi, Li, Yu and Zhang}]{Wang:13}
\bibinfo{author}{Wang, H.}, \bibinfo{author}{Shi, H.}, \bibinfo{author}{Li,
  Y.}, \bibinfo{author}{Yu, Y.}, \bibinfo{author}{Zhang, J.},
  \bibinfo{year}{2013}.
\newblock \bibinfo{title}{Seasonal variations in leaf capturing of particulate
  matter, surface wettability and micromorphology in urban tree species}.
\newblock \bibinfo{journal}{Frontiers of Environmental Science \& Engineering}
  \bibinfo{volume}{7}, \bibinfo{pages}{579--588}.
\newblock \DOIprefix\doi{10.1007/s11783-013-0524-1}.
\bibitem[{Zelikin et~al.(2017)Zelikin, Lokutsievskiy and Skopincev}]{ZLS:2017}
\bibinfo{author}{Zelikin, M.I.}, \bibinfo{author}{Lokutsievskiy, L.V.},
  \bibinfo{author}{Skopincev, S.V.}, \bibinfo{year}{2017}.
\newblock \bibinfo{title}{On optimal harvesting of a resource on a circle}.
\newblock \bibinfo{journal}{Mathematical Notes} \bibinfo{volume}{102},
  \bibinfo{pages}{521--532}.

\end{thebibliography}

\end{document}